\documentstyle[prl,aps]{revtex}
\input BoxedEPS.tex
\SetepsfEPSFSpecial
\HideDisplacementBoxes

\begin{document}

\twocolumn[\hsize\textwidth\columnwidth\hsize
         \csname @twocolumnfalse\endcsname
\title{Unambiguous evidence for  extended  s-wave pairing symmetry in 
hole-doped high-temperature superconductors} 

\author{Guo-meng 
Zhao$^{*}$} 

\address{ Department of Physics and Astronomy, California 
State University at Los Angeles, Los Angeles, CA 90032, USA}

\maketitle
\widetext
\begin{abstract}

We have analyzed data from angle resolved photoemission spectroscopy, 
Fourier 
transform scanning tunnelling spectroscopy, and low-temperature 
thermal conductivity for optimally doped 
Bi$_{2}$Sr$_{2}$CaCu$_{2}$O$_{8+y}$ in 
order to discriminate between d-wave and extended s-wave pairing 
symmetry.  The combined data are inconsistent with d-wave symmetry, 
but quantitatively consistent with extended s-wave symmetry with 
eight 
line nodes.  We also consistently explain all the phase-sensitive 
experiments.

\end{abstract}
\narrowtext
\vspace{0.3cm}

\narrowtext
]
 The phenomenon of superconductivity involves the pairing of 
electrons into Cooper pairs \cite{BCS}.  The internal wavefunction 
(gap function) of these Cooper pair obey a certain symmetry which 
reflects the underlying pairing mechanism.  It is known that 
conventional superconductors (e.g., Pb and Nb) possess an s-wave gap 
symmetry that reflects the phonon mediated electron-electron pairing 
\cite{BCS}.  On the other hand, the gap symmetry of high-temperature 
cuprate superconductors has been a topic of intense debate for over 
fifteen years.  Three symmetry contenders have been isotropic s-wave, 
d-wave and extended s-wave, as depicted in Fig.~1.  Both d-wave and 
extended s-wave have line nodes and change sign when a node is 
crossed.  A majority of experiments testing the symmetry (e.g., 
penetration depth, thermal conductivity, and specific heat 
measurements) have pointed to the existence of line nodes in the gap 
function \cite{Hardy,Jacobs,Lee,Chiao}.  Qualitatively, these 
experiments are consistent with both d-wave and extended s-wave gap 
functions.  
\begin{figure}[htb]
 \ForceWidth{7cm}
\centerline{\BoxedEPSF{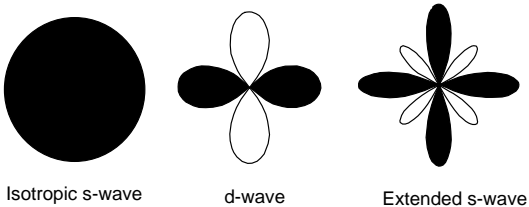}}	
	\vspace{1cm}
\caption[~]{Three allowed pairing symmetries appropriate for  
CuO$_{2}$ planes in the 
high-$T_{c}$ superconductors. Both d-wave and extended s-wave have 
line nodes and change signs when a node is 
crossed. }
\protect\label{Fig1}	
\end{figure}
There has been much experimental evidence for a d-wave 
symmetry of superconducting condensate (order parameter) 
for hole-doped cuprate superconductors.  In particular, 
phase-sensitive experiments based on planar Josephson tunneling 
\cite{Review} appear to provide compelling evidence for a d-wave 
order 
parameter symmetry.  Further, some angle resolved photoemission 
spectroscopy (ARPES) studies on nearly optimally doped 
Bi$_{2}$Sr$_{2}$CaCu$_{2}$O$_{8+y}$ (BSCCO) \cite{Shen,Ding1} show a 
very anisotropic gap that may be consistent with d-wave symmetry.  As 
a result, there is a widespread belief that the d-wave gap symmetry 
is 
now firmly established.  Nevertheless, there is also overwhelming 
evidence favoring an extended s-wave gap ($A_{1g}$ symmetry) 
\cite{Zhao,Brandow}.  This 
evidence includes data from phase-sensitive experiments based on 
out-of-plane Josephson tunneling \cite{Li,Klemn,Sun}, ARPES studies 
on heavily overdoped BSCCO \cite{Vob}, single-particle tunneling 
spectroscopy \cite{Mag}, Raman spectroscopy of heavily 
overdoped cuprates \cite{Kend}, Nonlinear Meissner effect 
\cite{Bha}, and inelastic neutron scattering \cite{ZhaoNeutron}.  
The measurements of the physical properties that are related to low 
energy quasiparticle excitations \cite{Hardy,Jacobs,Lee,Chiao} have 
definitively excluded a nodeless s-wave gap symmetry, but cannot 
distinguish between d-wave and extended s-wave unless one makes 
quantitative comparisons between theory and experiment.

Although there are more experiments favoring extended s-wave 
than d-wave gap symmetry \cite{Zhao,Brandow}, the extended 
s-wave evidence has been less well known and undervalued.  If the gap 
symmetry for hole-doped cuprate superconductors is extended s-wave, 
can we consistently explain all the phase-sensitive experiments? 

The next question is: If the intrinsic bulk gap symmetry is 
extended s-wave, can we definitively explain the ARPES data for 
BSCCO?   The ARPES data reported in the 1995 paper \cite{Ding} for a 
slightly overdoped BSCCO sample with $T_{c}$ = 87 K are consistent with 
an extended s-wave gap symmetry.  In contrast, the other ARPES data reported 
in the 1996 paper \cite{Ding1} for the same sample, which were 
measured by the same group with the same energy and momentum 
resolutions, are consistent with a simple d-wave gap function.  Now a question 
arises: Which ARPES data are more reliable.  One can easily check that 
both the Fermi-surface (FS) and superconducting gaps reported in 
the 1995 paper are in perfect agreement with those determined by 
independent and very precise Fourier transform scanning tunnelling 
spectroscopic (FT-STS) data \cite{Davis} (also see Fig.~2 below).  The 
FS reported in the 1995 paper is also in excellent agreement with that 
recently determined using the symmetrization method, the most reliable 
method to extract the FS from ARPES data \cite{Mesot}.  This indicates 
that the 1995 ARPES data are reliable because they are in perfect 
agreement with three independent sets of data, which are very precise 
and in perfect agreement with each other.  In contrast, the FS and 
superconducting gaps reported in the 1996 paper \cite{Ding1} are well 
off from those determined from the very precise FT-STS data (see 
Ref.~\cite{Davis}).  Moreover, the FS reported in the 1996 paper 
\cite{Ding1} disagrees significantly with that determined recently 
from the reliable symmetrization method \cite{Mesot}.  This implies 
that the 1996 ARPES data are not reliable because they are not 
consistent with other three independent sets of data, which are very 
precise and in perfect agreement with each other.
\begin{figure}[htb]
    \ForceWidth{7cm}
	\centerline{\BoxedEPSF{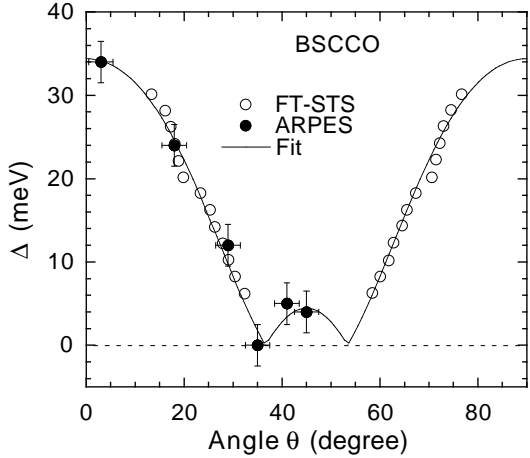}}
	\vspace{1cm}
	\caption[~]{The angle 
dependence of the superconducting gap 
$\Delta (\theta )$ in the Y quadrant for slightly overdoped  
Bi$_{2}$Sr$_{2}$CaCu$_{2}$O$_{8+y}$ crystals with $T_{c}$ 
= 86-87 K.  The gaps are extracted from ARPES data \cite{Ding} (solid 
circles) or from FT-STS data \cite{Davis} (open circles).  The solid 
line is the best fitted curve by Eq.~1. Here $\theta$ is the angle 
measured from 
the Cu-O bonding direction.  }
	\protect\label{Fig2}
\end{figure}
In Fig.~2 we plot the angle dependence of the superconducting gap 
$\Delta (\theta)$ in Y-quadrant for a slightly overdoped BSCCO with 
$T_{c}$ = 86-87 
K. The gaps are independently determined from ARPES \cite{Ding} and 
FT-STS 
studies \cite{Davis} on similar BSCCO crystals.  It 
is striking that two independent data sets are right on top of each 
other in the angle range accessible to both ARPES and FT-STS.  
The other two sets of the ARPES data \cite{Ding1,Ding2} 
are well off from the FT-STS data points \cite{Davis} due to the 
unreliable extraction of the FS (see the above discussion).  Only the 
gap near the antinodal direction is found to be the same in all 
the three measurements \cite{Ding1,Ding,Ding2}, suggesting that 
the error in the FS has little effect on the accuracy of the gap 
extraction along the antinodal direction.  Based on the reliable 1995 
ARPES data (see the above discussion), it is apparent that the gap at 
$\theta$ = 45$^{\circ}$ is finite rather than zero for this nearly 
optimally doped BSCCO.  This is further supported by another set of 
the ARPES data for a heavily overdoped BSCCO with $T_{c}$ = 60 K, 
which clearly shows that the gap at $\theta$ = 45$^{\circ}$ is about 
9$\pm$2 meV (Ref.~\cite{Vob}).  Recent ARPES data for Pb-doped BSCCO 
with $T_{c}$ = 70 K also indicate that the kink energies in the 
electronic dispersion along the ``nodal'' and antinodal directions 
differ by less than 6 meV (see Fig.~4b in Ref.~\cite{Lanzara}).  This 
implies that the gaps along the ``nodal'' and antinodal directions 
differ by less than 6 meV, and that the gap along the ``nodal'' 
direction is about 10 meV if the gap along the antinodal direction is 
about 16 meV.  These ARPES data thus suggest that the anisotropy 
between the gap at $\theta$ = 45$^{\circ}$ and at $\theta$ =0 
decreases with doping.  This conclusion is also consistent with Raman 
scattering data, which indicate that the gap difference between 
$\theta$ = 0 and 45$^{\circ}$ becomes negligibly small for heavily 
overdoped BSCCO and Tl$_2$Ba$_2$CuO$_{6+y}$ (Ref.~\cite{Kend}).

We can fit the combined data points in Fig.~2 by an extended s-wave 
gap 
function:
\begin{equation}\label{gap}
|\Delta (\theta )| = |\Delta (\cos 4\theta + s)+A\cos 8\theta|.
\end{equation}
 Here we include the next  harmonic term $\cos 8\theta$ of the 
 extended s-wave symmetry to account for 
 the high harmonic correction to the Fermi surface.  One can clearly  
see 
 that the fit is excellent with the fitting parameters: $\Delta$ 
=19.43(46) meV, $s$ = 0.874(22) and 
 $A$ = $-$2.01(41) meV.  From the fitted 
 curve, we find  that the line 
 nodes are located at $\theta_{n}$ = 36.7$^{\circ}$ and 
53.3$^{\circ}$ 
 in the first quadrant, and  the maximum 
 gap is 34.3 meV. Since the intrinsic bulk 
 maximum gap for the optimally doped BSCCO is 33-34 meV, as seen from 
intrinsic tunneling 
 spectroscopy \cite{Krasnov}, we conclude that the top CuO$_{2}$ 
layer 
 of this slightly overdoped BSCCO is optimally doped.
 
 Because the ARPES determined gap near the antinodal direction is 
very accurate with an uncertainty of $\pm$1.5 meV (Ref.~\cite{Ding}),  we can fit only the FT-STS data with a constraint that the fitted 
curve is exactly through one ARPES data point near the antinodal 
direction (the first ARPES data point in Fig.~2).  Such a fit leads to 
almost the same fitting parameters as the fit that includes all the 
ARPES data points.

The FT-STS data points alone could  be also consistent with a d-wave 
gap 
function:
 \begin{equation}\label{dgap}
|\Delta (\theta )| = |\Delta_{M} [(1-B)\cos 2\theta +B\cos 6\theta ]|.
\end{equation}
Here  $\cos 6\theta$ is the next  harmonic term  
 of the d-wave gap function. McElroy {\em et al. } \cite{Davis} fit 
their FT-STS 
 data by Eq.~\ref{dgap}.  The best fit leads to the fitting 
 parameters: $\Delta_{M} $ = 39.3 meV and $B$ = 
 0.182.  Although the d-wave fit is also good, the fitted maximum gap 
(39.3 meV) 
 is well above the value (34 $\pm$1.5 meV) found from ARPES 
 (Ref.~\cite{Ding}).  Because the upper limit of the antinodal gap is 
 bounded by the peak position of the spectrum, which is 35 meV below 
 the Femi level \cite{Ding}, the fitted maximum gap (39.3 meV) is 
 unphysical.

 In order to definitively exclude the d-wave gap symmetry, one could 
 quantitatively compare both d-wave and extended s-wave predictions 
 with other experiments.  One such prediction is the linear slope of 
 the low-temperature electronic thermal conductivity, 
 $\kappa_{\circ}/T$, which is directly related to the Fermi velocity 
 $v_{F}$ and momentum $k_{F}$ in the nodal directions, and the slope 
$S 
 = d\Delta (\theta$)/d$\theta$ at nodes.  The former two quantities 
can 
 be obtained from ARPES data while the latter one can be readily 
 calculated from the gap function deduced from a fit.  The residual 
 thermal conduction is due to a fluid of zero-energy quasiparticles 
 induced by the pair-breaking effect of impurity scattering near the 
 nodes in the gap.  Calculations for the heat transport by nodal 
 quasiparticles in two dimensions give a general expression \cite{Lee1}
\begin{equation}\label{TM}
\frac{\kappa_{\circ}}{T}=N\frac{k_{B}^{2}}{3\hbar}\frac{n}{d}(\frac{v_{F}}{v_{2}}+\frac{v_{2}}{v_{F}}),
\end{equation}
where $n/d$ is the stacking density of CuO$_{2}$ planes, $v_{2} = 
S/(\hbar k_{F})$, and $N$ is the number of nodes per quadrant.  
Remarkably, theory \cite{Lee1} predicts that $\kappa_{\circ}/T$ is 
independent of
impurity concentration and that  Eq.~\ref{TM} is still valid 
even when vertex and Fermi-liquid corrections are taken 
into account.  This makes thermal conductivity a very robust probe of 
the nodal quasiparticle spectrum in anisotropic superconductors.

A detailed ARPES study on BSCCO \cite{Kam} shows that $v_{F}$ depends 
on angle $\theta$ being maximum at $\theta$ = 45$^{\circ}$ while 
$k_{F}$ only has a weak angle dependence ($k_{F}$=0.74 \AA$^{-1}$ 
near 
$\theta$ = 45$^{\circ}$).  The $v_{F} (\theta)$ can be evaluated from 
a linear fit to the band dispersions (energy distribution curves) 
above a kink energy (about 50 meV below the Fermi level).  From the 
middle column of Fig.~4 in Ref.~\cite{Kam}, we estimate that $\hbar 
v_{F}$=1.40 eV\AA ~for $\theta$ = 45$^{\circ}$ and $\hbar v_{F}$=1.16 
eV\AA ~for $\theta$ = 38.5$^{\circ}$.  If we linearly extrapolate 
$v_{F}$ with $\theta$, we obtain $\hbar v_{F}$ = 1.09 eV\AA ~at an 
extended s-wave gap nodal angle of $\theta_{n} $ = 36.7$^{\circ}$ 
deduced above for the optimally doped  top CuO$_{2}$ layer of 
BSCCO.  Similarly, we estimate $\hbar v_{F}$ = 0.69 eV\AA ~for $\theta$ 
= 25.8$^{\circ}$, which is one of the nodal directions of the extended 
s-wave gap function deduced for slightly overdoped 
YBa$_{2}$Cu$_{3}$O$_{7-y}$ (see below).  It is interesting to note 
that the value of $v_{F}$ (1.05$\times$10$^{5}$ m/s) at the nodal 
directions of YBa$_{2}$Cu$_{3}$O$_{7-y}$ (YBCO) estimated from ARPES is very close to the value, 
(1.2$\pm$0.2)$\times$10$^{5}$ m/s, estimated from the field dependence 
of the in-plane magnetic penetration depth at low temperatures 
\cite{Carrington2001}.
\widetext
\begin{table}[htb]
 \caption[~]{The calculated values of $\kappa_{\circ}/T$ for the 
 optimally doped BSCCO in terms of the d-wave and extended s-wave gap 
 functions deduced from the best fits to the ARPES and FT-STS data 
 for the optimally doped top layer of BSCCO (see text).  The 
experimental value 
 of $\kappa_{\circ}/T$ for optimally doped BSCCO is about
 0.2 mW/K$^{2}$cm (see text).  Here $d/n$ is the average 
 separation between CuO$_{2}$ planes stacked along the c axis, 
 $\theta_{n}$ is the angle of the nodal direction, $S = d\Delta 
 (\theta$)/d$\theta$ is the slope at the node and $N$ is the number 
of 
 nodes per quadrant.  }
\begin{center}
\begin{tabular}{lccccccc}
&$d/n$ (\AA)&$\theta_{n}$&S/N (meV) &$\hbar v_{F}$(eV\AA) &$k_{F}$ 
(\AA$^{-1}$) &$v_{F}/v_{2}$& $\kappa_{\circ}/T$ (mW/K$^{2}$cm)\\
\hline
d-wave&7.72&45$^{\circ}$&21.38&1.40&0.74&  48.5&  0.378\\
Extended s-wave &7.72&36.7$^{\circ}$&28.6&1.09&0.74& 28.2& 0.220\\ 
	\end{tabular}
	\end{center}
	\end{table}
	\narrowtext

With the values of $\hbar v_{F}$ and $k_{F}$, we can now 
calculate $\kappa_{\circ}/T$ for the d-wave and the 
extended s-wave gap functions using Eq.~\ref{TM}.  The calculated 
values of $\kappa_{\circ}/T$ are shown in the last 
column of Table 1.  Since the top CuO$_{2}$ layer of this slightly 
overdoped BSCCO is optimally doped, as discussed above, we should 
compare the calculated values of $\kappa_{\circ}/T$ with the measured 
one for an optimally doped BSCCO, which is not available. 
Fortunately, 
it is known that the value of $\kappa_{\circ}/T$ for slightly 
underdoped  cuprates is 
slightly higher than that for  slightly overdoped cuprates.  This can 
be 
seen clearly from the YBCO system: $\kappa_{\circ}/T$= 0.17$\pm$0.01 
mW/K$^{2}$cm for slightly underdoped YBa$_{2}$Cu$_{3}$O$_{6.90}$ 
(Ref.~\cite{Taillefer}) and $\kappa_{\circ}/T$= 0.12$\pm$0.02 
mW/K$^{2}$cm for  overdoped YBa$_{2}$Cu$_{3}$O$_{7.0}$ 
\cite{Chiao}.  By analogy,  we should take the experimental value of 
$\kappa_{\circ}/T$ to be about 0.2 mW/K$^{2}$cm for optimally doped 
BSCCO, which is slightly  larger than that for slightly overdoped 
BSCCO (0.15 mW/K$^{2}$cm \cite{Chiao})

From Table 1, 
one can clearly see that the predicted value of $\kappa_{\circ}/T$ 
from the d-wave gap function is larger than the experimental value by 
a factor of 2 while this value from an extended s-wave gap model is 
within 10$\%$ of the experimental value.
This  indicates that the gap symmetry for the optimally doped 
BSCCO is not d-wave but extended s-wave.

For optimally doped  YBa$_{2}$Cu$_{3}$O$_{6.93}$, the measured 
$\kappa_{\circ}/T$ is about 0.17 mW/K$^{2}$cm (Ref.~\cite{Taillefer}) 
and $\Delta_{M}$ $\simeq$ 30 meV.  Taking 
the d-wave gap form of Eq.~\ref{dgap} with $B$ = 0.182 and 
$\Delta_{M}$ = 30 meV, we have $\kappa_{\circ}/T$ = 0.65 mW/K$^{2}$cm, 
which is larger than the measured one by a factor of 4.

On the other hand, an 
extended s-wave gap function $\Delta (\theta ) = 24.5 (\cos 4\theta + 
0.225)$ meV has been deduced from a single-particle tunneling 
spectrum for slightly overdoped YBCO \cite{Zhao}.  This gap function has 
line nodes located at $\theta_{n}$ = 25.8$^{\circ}$ and 64.2$^{\circ}$ 
in the first quadrant.  With ~$\hbar v_{F}$ = 0.69 eV\AA~(see above), 
$k_{F}$= 0.74 \AA$^{-1}$, and $S$ = 47.7 meV (evaluated from the 
extended s-wave gap function), we calculate $\kappa_{\circ}/T$ = 0.12 
mW/K$^{2}$cm, in quantitative agreement with the measured one 
(0.14$\pm$0.02 mW/K$^{2}$cm) \cite{Chiao}. 
 
The above quantitative data analyses unambiguously show that the  
gap symmetry for optimally doped cuprates is extended s-wave.  How 
could this conclusion be compatible with all the phase-sensitive 
experiments?  This issue can be resolved if we consider the fact that 
\newpage
\noindent 
there are two types of charge carriers; one is 
intersite bipolarons of oxygen holes and another is Fermi-liquid type 
with a large Fermi surface \cite{Muller}.  The Fermi-liquid component 
is nearly absent for the hole doping $p$ $<$ 0.1, and increases 
monotonically with doping for $p$ $>$ 0.1 (Ref.~\cite{Muller}).  
Further, it is shown that the Bose-Einstein condensate of bipolarons 
has d-wave symmetry \cite{Alex}.  Because the interfaces of 
grain-boundary junctions consist of underdoped cuprates \cite{Bet}, 
the d-wave component of Bose-Einstein condensate of bipolarons is 
dominant at the surface, in agreement with phase-sensitive experiments 
based on grain-boundary Josephson junctions \cite{Review}.  It was 
also shown that the surface layer of a cuprate crystal is underdoped 
when it is contacted with a normal metal \cite{Mann}.  This can 
explain the observation of dominant d-wave component in the corner 
SQUID experiments \cite{Wollman,Mathai}.  The extended s-wave symmetry 
for the Fermi-liquid component can naturally account for 
phase-sensitive experiments based on out-of-plane Josephson tunneling 
\cite{Sun}.

 In summary, we have analyzed data from angle resolved photoemission 
spectroscopy, Fourier 
transform scanning tunnelling spectroscopy, and low-temperature 
thermal conductivity for optimally doped 
Bi$_{2}$Sr$_{2}$CaCu$_{2}$O$_{8+y}$ in 
order to discriminate between d-wave and extended s-wave pairing 
symmetry.  The combined data are inconsistent with d-wave symmetry, 
but quantitatively consistent with extended s-wave symmetry with 
eight 
line nodes. 
 ~\\
~\\
$^{*}$Correspondence should be addressed to gzhao2@calstatela.edu


\begin{thebibliography}{99}
\bibliographystyle{prsty} 
\bibitem{BCS} 
J.  Bardeen, L.  N.  Cooper, and J.  R. Schrieffer,  Phys.  Rev.  
{\bf 108}, 1175 (1957).

 \bibitem{Hardy}W. N. Hardy, D. A. Bonn, D. C. Morgan, Ruixing Liang, 
and K. Zhang, Phys. Rev.
Lett. \textbf{70},
3999 (1993).  

\bibitem{Jacobs}T. Jacobs, S. Sridhar, Q. Li, G. D. Gu, and N. 
Koshizuka, Phys. Rev. 
Lett. \textbf{75}, 4516 (1995).

\bibitem{Lee}S.-F. Lee, D. C. Morgan, R. J. Ormeno, D. Broun, R. A. 
Doyle, J. R. Waldram, and K. Kadowaki, Phys. Rev. 
Lett. \textbf{77}, 735 (1996).

\bibitem{Chiao}M. Chiao, R. W. Hill, C. Lupien, L. Taillefer, P. 
Lambert, R. Gagnon, and P. Fournier, Phys. Rev. 
B \textbf{62}, 3554 (2000).

\bibitem{Review} C.  C. Tsuei and  J.  R.  Kirtley, Rev.  Mod.  
Phys. {\bf 72}, 969 (2000).

\bibitem{Shen} Z.-X. Shen et al.,   Phys.  Rev.  Lett.  
{\bf 70}, 1553 (1993).


\bibitem{Ding1} H. Ding et al.,  Phys.  Rev.  B.  {\bf 54}, R9678 
(1996).  
\bibitem{Zhao} G.  M. Zhao,   Phys.  Rev.  B {\bf 64}, 024503 (2001).
\bibitem{Brandow} B. H. Brandow, Phys.  Rev.  B {\bf 65}, 054503 
(2002).

\bibitem{Li}Q. Li, Y. N. Tsay, M. Suenaga, R. A. Klemm, G. D. Gu, and 
N. 
Koshizuka, Phys. Rev. 
Lett. \textbf{83}, 4160 (1999). 

\bibitem{Klemn} A.  Bille, R.  A.  Klemm, and K. Scharnberg,   
Phys.  Rev.  B {\bf 64}, 174507 (2001).

\bibitem{Sun}A. G. Sun, D. A. Gajewski, M. B. Maple, and R. C. Dynes, 
Phys. Rev. 
Lett. \textbf{72}, 2267 (1994).
 
 
\bibitem{Vob}I. Vobornik, R. Gatt, T. Schmauder, B. Frazer, R. J. 
Kelley, C. Kendziora, M. Grioni, M. Onellion, and G. Margaritondo, 
Physica C \textbf{317-318}, 589 (1999).

\bibitem{Mag}I. Maggio-Aprile, Ch. Renner, A. Erb, E. Walker, and O. 
Fischer, Phys. Rev. Lett. \textbf{75}, 2754 (1995).

\bibitem{Kend}C. Kendziora, R. J. Kelley, and M. Onellion, Phys. Rev. 
Lett. \textbf{77}, 727 (1996).

\bibitem{Bha}A. Bhattacharya, I. Zutic,  O. T. Valls, A. M. Goldman, 
U. 
Welp, and B. Veal, Phys. Rev. 
Lett. \textbf{82}, 3132 (1999).

\bibitem{ZhaoNeutron} G. M. Zhao,  cond-mat/0302566.  

\bibitem{Ding}H. Ding {\em et al.}, Phys. Rev.  Lett. \textbf{74}, 
2784 (1995).



 \bibitem{Davis} K.  McElroy {\em et al.,} Nature (London) {\bf 
422}, 592 (2003).  

\bibitem{Mesot}J.  Mesot {\em et al.}, Phys.  Rev.  B {\bf 63}, 
224516 (2001).

\bibitem{Ding2}J.  Mesot {\em et al.}, Phys.  
Rev.  Lett.  {\bf 83}, 840 (1999).



\bibitem{Lanzara}A.  Lanzara {\em et al.}, Nature (London) {\bf 412}, 510 (2001).

\bibitem{Krasnov}V.  M.  Krasnov, A.  Yurgens, D.  Winkler, P. 
Delsing, 
and T. Claeson,   Phys.  rev.  Lett.  {\bf 84}, 5860 (2000).

\bibitem{Lee1}A.  C.  Durst and P.  A. Lee,   Phys.  Rev.  B {\bf 
62}, 1270 (2000).
 
\bibitem{Kam} A.  Kaminski {\em et al.,} Phys.  Rev.  Lett.  
\textbf{86}, 1070  (2001).

\bibitem{Carrington2001} A. Carrington, F. Manzano, R. Prozorov, R. 
W. Giannetta, N. Kameda, and T. Tamegai, Phys. Rev. Lett. 
\textbf{86}, 1074 (2001).

\bibitem{Taillefer}
L. Taillefer, B. Lussier, R. Gagnon,  K. Behnia, and H. Aubin, Phys.  
Rev.  Lett.  \textbf{79}, 483 (1997).

 
 \bibitem{Muller}K. A.  M\"uller, Guo-meng Zhao, K. Conder, and H.  
Keller, 
 J.  Phys.: Condens.  Matter, {\bf 10}, L291 (1998).




\bibitem{Alex} A. S. Alexandrov,   Physica C \textbf{305}, 46 (1998).

\bibitem{Bet} J.  Betouras and R. Joynt,  Physica C {\bf 250}, 256 
(1995).
\bibitem{Mann}J. Mannhart and H. Hilgenkamp, Physica C 
\textbf{317-318}, 
383 (1999).
\bibitem{Wollman} D.  A.  Wollman, D.  J.  Van Harlingen, J.  
Giapintzakis, 
and D.  M.  Ginsberg, Phys.  Rev.  Lett.  {\bf 74}.  797 (1995).  
\bibitem{Mathai}A.  Mathai, Y.  Gim, R.  C.  Black, A.  Amar, and F.  
C.  Wellstood, Phys.  Rev.  Lett.  {\bf 74}, 4523 (1995).
\end{thebibliography}
\end{document}